\documentclass[aps,prl,twocolumn,superscriptaddress]{revtex4}

\usepackage{epsfig}
\def\beq{\begin{equation}}
\def\eeq#1{\label{#1}\end{equation}}
\def\eeqn{\end{equation}}
\def\beqa{\begin{eqnarray}}
\def\eeqa#1{\label{#1}\end{eqnarray}}
\def\eeqan{\end{eqnarray}}

\def\to{\rightarrow}

\newcommand\iden{\leavevmode\hbox{\small1\normalsize\kern-.33em1}}

\def\W3{W_H^3}

\begin{document}


\title{Neutrino Telescopes as a Direct Probe of Supersymmetry Breaking}
\author{Ivone~Albuquerque}
\affiliation{Department of Astronomy and Space Sciences Laboratory, University
of California, Berkeley, CA~94720}
\author{Gustavo Burdman} 
\affiliation{Theory Group, Lawrence Berkeley National Laboratory, Berkeley, CA~94720}
\affiliation{Department of Physics, University
of California, Berkeley, CA~94720}
\author{Z. Chacko}
\affiliation{Theory Group, Lawrence Berkeley National Laboratory, Berkeley, CA~94720}
\affiliation{Department of Physics, University
of California, Berkeley, CA~94720}


\begin{abstract}

We consider supersymmetric models where the scale of supersymmetry breaking lies between 5 $\times
10^6$ GeV and 5 $\times 10^8$ GeV. In this class of theories, which includes models of gauge
mediated supersymmetry breaking, the lightest supersymmetric particle is the gravitino. The next
to lightest supersymmetric particle is typically a long lived charged slepton with a lifetime
between a microsecond and a second, depending on its mass. Collisions of high energy neutrinos
with nucleons in the earth can result in the production of a pair of these sleptons. Their very
high boost means they typically decay outside the earth. We investigate the production of these
particles by the diffuse flux of high energy neutrinos, and the potential for their observation in
large ice or water Cerenkov detectors. The relatively small cross-section for the production of
supersymmetric particles is partially 
compensated for by the very long range of heavy particles. The signal
in the detector consists of two parallel charged tracks emerging from the earth about 
100 meters apart, with very little background. A detailed calculation using the Waxman-Bahcall limit
on the neutrino flux and realistic spectra shows that km$^3$ experiments 
could see as many as 4 events a year.
We conclude that neutrino telescopes will complement 
collider searches in the determination of the supersymmetry breaking scale, and  
may even give the first evidence for supersymmetry at the weak scale.

\end{abstract}

\maketitle

{\it Introduction ---} 
The origin of the radiative stability of the weak scale is one of the most important questions in
particle physics today. A natural answer requires new physics at the TeV scale. Among the
candidate theories, weak scale supersymmetry remains the most attractive scenario. Although this
is in no small measure due to its theoretical appeal (it is a simple and natural extension of the
usual space-time symmetries), it is also favored by data from electroweak observables. These point
to a weakly coupled Higgs sector, one without significant deviations from the standard model in
regard to electroweak precision observables. However, supersymmetry must be broken since the
superpartners have not yet been observed. The supersymmetric spectrum is determined
by the supersymmetry breaking mechanism.\\
\indent 
Supersymmetric models typically have a symmetry, called R-parity, which ensures that the
Lightest Supersymmetric Particle, the `LSP', is stable. Which of the supersymmetric particles
is the LSP? This is determined by the scale of supersymmetry breaking, which we denote by
$\sqrt{F}$, and which can lie anywhere between $10^3$ GeV and $10^{12}$GeV. If supersymmetry
is broken at high scales such that $\sqrt{F}$ is larger than $10^{10}$GeV the LSP is typically
the neutralino. If however supersymmetry is broken at lower scales, $\sqrt{F} < 10^{10}$GeV, 
the LSP is typically the gravitino.
In models where the LSP is the gravitino, the Next to Lightest Supersymmetric Particle
(NLSP) tends to be a charged slepton, typically
the right-handed stau. Since the NLSP decays to gravitinos through interactions that are 
suppressed by powers of $\sqrt{F}$, if the supersymmetry breaking scale is high 
its lifetime can be quite large. In gauge-mediated SUSY breaking, for instance, we have
\beq 
 c\tau = \left(\frac{\sqrt{F}}{10^7{\rm~GeV}}\right)^4\, 
\left(\frac{100~{\rm GeV}}{m_{\tilde\tau_R}}
\right)^5\,10~{\rm km}~,
\label{ctau}
\eeqn
where ${m_{\tilde\tau_R}}$ is the stau mass. Thus, for $\sqrt{F} \ge 10^7$GeV if 
these NLSPs were
to be produced by very high energy collisions they could travel very long distances before
decaying. In the last several years many interesting and realistic scenarios 
have been proposed in
which the scale of supersymmetry breaking $\sqrt{F}$ is low and could lie between 
$5 \times 10^6$GeV and
about $5 \times 10^8$GeV. These include models of gauge mediation{\cite{GMSB}}, and gauge and
Yukawa mediation{\cite{YMSB}}, warped higher dimensional models in which 
supersymmetry is broken on
an infrared brane and therefore the scale $\sqrt{F}$ has been warped down {\cite{warp}}, theory
space realizations of higher dimensional models {\cite{deconstruct}} and models of supersoft
supersymmetry breaking which are characterized by Dirac gauginos{\cite{Dirac}}. In all these
classes of models the NLSP is typically a right handed stau.

The existence of diffuse fluxes of high energy neutrinos, possibly associated with the production
of cosmic rays, has been widely discussed in the literature. Collisions of these high energy
neutrinos with nucleons in the earth at energies above threshold for supersymmetric production
frequently result in the production of a pair of supersymmetric particles, which promptly decay
into NLSPs. These typically have a high boost $\gamma_{\rm NLSP}\simeq 1000$ or larger and
therefore will not decay inside the earth provided the supersymmetry breaking scale $\sqrt{F} >
10^7$GeV. For $5 \times 10^6 < \sqrt{F} < 10^7$GeV a significant fraction of the decays will occur
inside the earth. Since the NLSP is charged, its upward going tracks could in principle be
detected in large ice or water Cerenkov detectors, such as ICECUBE~\cite{icecube}. This is in
analogy with the standard model charged current interaction giving muons, the primary signal in
neutrino telescopes.

Naively, one would expect that the event rates for NLSPs -which typically have masses in the
$100$~GeV range- are negligible when compared to 
those for muons, since their production cross section
must be considerably smaller than that of the standard model interactions. The
reason is that the SM interactions come primarily
from very small values of $x$, the parton momentum fraction, whereas the supersymmetric
process is limited to $x> m^2/2M_PE_\nu$, with $m$ given essentially by the sum of the
produced supersymmetric particles. This naive expectation is however, misleading. The crucial
observation is that the range of a slepton is much larger than that of a muon, since energy
loss due to radiation sets in at much higher energies. In neutrino telescopes, muon events
must be produced either right outside the detector or in it, since the muon range for the
energies of interest is in the few to tens of kilometers.  Thus, most of the upgoing CC
events produced in the earth are lost. On the other hand, the range of NLSPs is typically in
the hundreds to thousands of kilometers. Then, unlike for the muon~\cite{als}, a significant 
fraction of the NLSPs produced will range into the detector.

In what follows we compute the number of NLSP events. For this purpose we calculate the
supersymmetric production cross sections and the NLSP range.

\begin{figure}
\centering
\epsfig{file=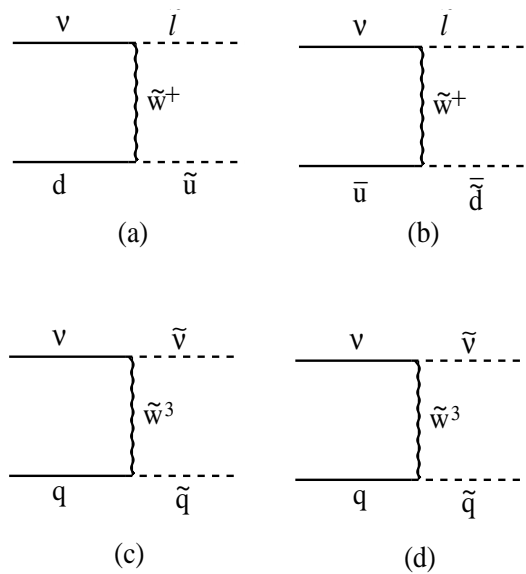,width=7cm,height=8cm,angle=0}
\caption{
Feynman diagrams for supersymmetric particle production in $\nu N$ collisions 
Charged current (chargino) interactions: (a) Left-left interaction requiring
the insertion of the gaugino mass in the t-channel line. (b) Left-right interaction.  
Neutral current: (c), (d). 
There are analogous diagrams for anti-neutrinos as well as for strange and charm initial 
quarks. 
}
\label{feynman}
\end{figure}

{\it The SUSY Cross Section---}
In the scenarios under consideration, every $\nu N$ interaction producing
supersymmetric particles will result in a pair of NLSPs, which have a 
very long lifetime. In 
what follows, we will assume this lifetime to be large 
enough so that NLSPs do not decay in the earth. 
For simplicity, we also neglect mixing with Higgsinos in the gaugino sector.
The dominant process is analogous to the SM CC interactions and corresponds to the 
t-channel exchange of charginos producing $\nu N\to \tilde\ell~\tilde q$, as shown in 
Figure~\ref{feynman}(a) and (b). The neutrino, always produced left-handed by the 
weak interactions, can interact either with a left-handed down-type quark (a), or with a 
right-handed up-type quark (b). 
This results in the partonic cross sections:
\beqa
\frac{d\sigma^{\rm (a)}}{dt} &=& \frac{\pi\alpha}{2\,\sin^4\theta_W}\,
\frac{M^2_{\tilde w}}{s\,(t-M^2_{\tilde w})^2}
\label{llcc}\\
\frac{d\sigma^{\rm (b)}}{dt} &=& \frac{\pi\alpha}{2\,\sin^4\theta_W}\,
\frac{(tu-m_{\tilde\ell_L}^2\,m_{\tilde q}^2)}{s^2\,(t-M^2_{\tilde w})^2}
\label{lrcc},
\eeqan
where $s,t$ and $u$ are the usual Mandelstam variables, and 
$M_{\tilde w}$, $m_{\tilde\ell_L}$ and $m_{\tilde q}$ are the chargino, 
the left-handed slepton and the squark masses respectively. 
The left-handed slepton and the squark decay promptly to the lighter ``right-handed'' slepton 
plus 
non-supersymmetric particles. 
We also include the subdominant neutralino exchange, Figure~\ref{feynman}~(c)-(d). 
\begin{figure}[t]
\centering
\epsfig{file=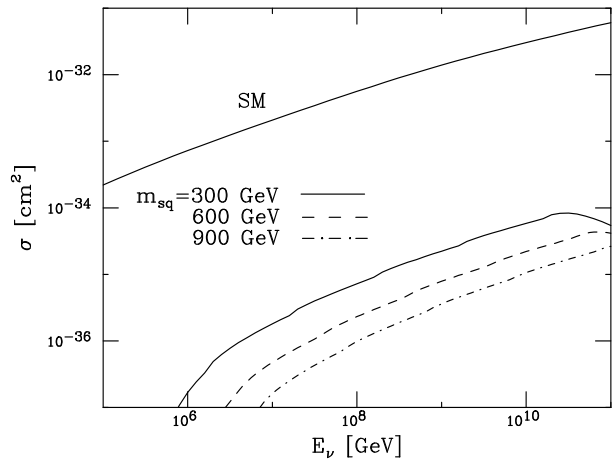,width=6cm,height=8cm,angle=90}
\caption{$\nu N$ cross sections vs. the energy of the incident neutrino.
The curves correspond to $m_{\tilde\ell_L}=250~$GeV, $m_{\tilde w}=250$GeV; and 
for squark masses  
$m_{\tilde q}=300$~GeV (solid) ,~$600$~GeV (dashed) and $900~$GeV (dot-dashed).
The top curve corresponds to the SM charged current interactions. 
}
\label{fig1}
\end{figure}   
 We take $m_{\tilde w}=250$~GeV, $m_{\tilde\ell_L}=250$~GeV and three values for the   
squark masses: $m_{\tilde q}= 300,~600$ and $900$~GeV. These are very representative 
values in the scenarios under consideration. Typically, the $\tilde\tau_R$ is the NLSP, being
heavier only than the ultra-light and very weakly coupled gravitino. 
Charginos and neutralinos tend to be heavier since
they also feel the $SU(2)_L$ interactions. Finally, squarks are heavier still since their masses
affected by the strong interactions. 
In Figure~\ref{fig1} we plot the cross sections for supersymmetric 
production in $\nu N$ interactions
as a function of the neutrino energy. Also plotted for comparison is the SM charged current 
cross section.
As advertised earlier, the SUSY cross sections are still suppressed with respect to  the SM, 
even when well above threshold. 

{\it The NLSP Range---}
Once produced by the $\nu N$ interactions in the earth, the NLSP pair should 
range into the detector, just as the muons produced by CC events~\cite{als}. 
Charged particles lose energy due to ionization processes as well as through radiation.
The average energy loss is given by~\cite{pdg}
\beq
-\frac{dE}{dx} = a(\beta\gamma) + c(\beta\gamma)~\beta\gamma~,
\label{eloss}
\eeq
where $a$ and $c$ characterize the ionization and radiation losses respectively, and 
are slowly varying functions of the energy. 
The ionization loss can be approximated by 
\beq
a(\beta\gamma)\simeq 0.08~\frac{\rm MeV~cm^2}{\rm gr}\,\left(17 + 2\ln\beta\gamma\right),
\label{ae}
\eeq
and is rather independent of the particle mass.  On the other hand, 
assuming $c(\beta\gamma)\simeq {\rm const.}$, the radiative energy 
loss can be written as $c~\beta\gamma = (b_m) E$. Thus 
$b_\mu m_\mu = b_{\tilde\tau_R}m_{\tilde\tau_R}$, and the radiative energy loss for the
NLSPs scales inversely with the mass. This results in a much larger range for the 
NLSP as compared to the muon. 
Current bounds on $m_{\tilde\tau_R}$ are just above $100$~GeV. As a reference value we 
take $m_{\tilde\tau_R}=150$~GeV.
Therefore, NLSPs produced hundreds, even thousands 
of kilometers away are within range of the detector.
This is to be contrasted with the fact that muons must be produced
at distances not larger than tens of kilometers 
from the detector in order to be observed. 
As we will see, this will somewhat compensate the suppression 
of the SUSY cross sections observed earlier.

{\it Signals in Neutrino Telescopes---} 
In order to compute the event rates in neutrino telescopes, we need to know the 
incoming neutrino flux. The presence of cosmic neutrinos is expected on the basis of the 
existence of high energy cosmic rays. Several estimates of the neutrino flux  are available 
in the 
literature. In most cases, it is expected that km$^3$ neutrino telescopes
will measure this flux. Here, in order to present projections for the 
number of observed 
SUSY events, we make use of  the Waxman-Bahcall (WB) limit~\cite{wb} as an estimate of the 
cosmic neutrino flux. We consider an initial flux containing both $\nu_\mu$ and $\nu_e$~(in a 
$2:1$ ratio). Since the 
initial interactions (see Figure~\ref{feynman}) produce $\tilde\ell_L$  and these 
are nearly degenerate 
in flavor, the flavor of the initial neutrino does not affect our results. 
For the same reason, the possibility of large mixing in the neutrino flux is also innocuous here. 
In order to correctly take into account the propagation of neutrinos and the NLSP
$\tilde\ell_R$ through the earth, we make use of a model of the earth density profile  as detailed
in Reference~\cite{gqrs}. 

\begin{figure}
%
\begin{center}
\epsfig{file=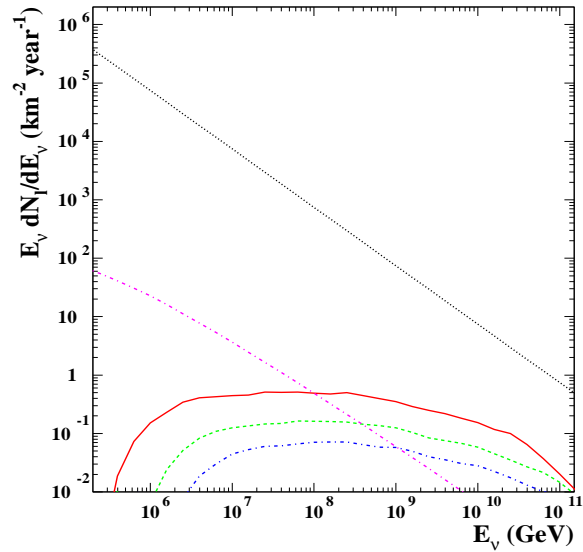,width=7.2cm,height=8cm,angle=0}
\caption{Energy distribution of $\tilde\tau_R$ pair events per km$^2$, per year. 
From top to bottom: $m_{\tilde q}=300$, ~$600$ and $900$~GeV. Here, $m_{\tilde\tau_R=150}$~GeV and
$m_{\tilde w}=250$~GeV. 
Also shown are the neutrino flux at earth and the $\mu$ flux 
through the detector. In all cases we make use of the WB limit for the neutrino flux. 
}
\label{flux}
\end{center}
\end{figure}
In Figure~\ref{flux} we show the energy distribution for the NLSP pair events for three
choices of squark masses: $300$~GeV, $600$~GeV and $900$~GeV. Also shown are the neutrino 
flux at earth
in the WB limit, as well as the energy distribution of upgoing $\mu$'s. 
We see that, even for the heavier squarks, it is possible to obtain observable event rates. 
In Table~\ref{table} we show the event rates for $\tilde\ell_R$ pair production
per year and per km$^2$. The rates are given for the WB flux as well as for the 
Mannheim-Protheroe-Rachen (MPR) flux~\cite{mpr}, both for optically thin sources.  
For comparison, we also show the rates of upgoing muons. Thus, km$^3$ Cerenkov detectors such as 
ICECUBE, appear to be sensitive to most of the parameter space of interest in scenarios with 
a relatively long lived NLSP.

\begin{table}
\begin{center}
\caption{Number of events per km$^2$ per year for the WB and MPR fluxes. The first column refers to 
upgoing muons. The last three columns correspond to upgoing NLSP pair events, 
for three different choices of squark masses: $300$~GeV, $600$~GeV and $900$~GeV.
The number of muon events are given for energies
above threshold for production of a $250$~GeV $\tilde\ell_L$ plus a $300$~GeV
squark, ie, $1.6 \times 10^5$ GeV.}
\begin{ruledtabular}
\begin{tabular}{l|cccc}
 & $\mu$ & $m_{\tilde{q}}=300$~GeV & $600$~GeV & $900$~GeV \\
\hline
WB & 106 & 4 & 1 & 0.5 \\
MPR & 1085 & 10 & 3 & 1 
\label{table}
\end{tabular}
\end{ruledtabular}
\end{center}
\end{table}

Since the NLSPs are produced in pairs very far from the detector and with a very large boost, 
typical signal events consist of {\rm two} tracks separated by 
$\delta R\simeq L~\theta$, 
with $L$ the distance to the production point ($\simeq 100-1000$~Km) and 
$\theta\simeq p_{\rm SUSY}^{\rm CM}/p_{\rm boost}\simeq 10^{-3}-10^{-4}$.
If we consider $L$ to be of the order of the NLSP range, then in the linear regime 
$\delta R\simeq {\rm constant} \simeq 100{\rm m}$. As we have seen in the discussion 
following eqn.~(\ref{ae}),  
for very high energies
the range grows logarithmically with energy, leading to somewhat smaller values 
of $\delta R\simeq (20-40){\rm m}$. The track separation is then mildly 
sensitive to the stau injection energy~\cite{abc2}.   
Then, most NLSP events would consist of two parallel but well separated tracks, 
and are therefore expected to be very distinctive and different from backgrounds. 
In addition, we see from Figure~\ref{flux} that the signal events have a harder spectrum 
than the $\mu$ background. 

{\it Conclusions ---}
We have shown for the first time that neutrino telescopes are potentially sensitive to the
relatively long-lived charged NLSPs which are present in a wide variety of models of
supersymmetry breaking.  The event rates shown in Table~\ref{table} are already encouraging
for experimental facilities that are been built, such as ICECUBE. The region of the
supersymmetry breaking parameter space that is available to neutrino telescopes is determined,
on the one hand, by the twin requirements that the NLSP lifetime be long enough to give a
signal ($\sqrt{F}\agt 5 \times 10^6$~GeV), but not be so long as to disturb Big Bang
Nucleosynthesis ($\sqrt{F}\alt 5 \times 10^8$~GeV). Thus the observation of NLSP
events at neutrino telescopes will constitute a direct probe of the scale of supersymmetry
breaking.  This is to be compared with the potential observation of these NLSPs at the Large
Hadron Collider (LHC), where for this range of $\sqrt{F}$,  the NLSP decays outside the
detector and is seen through its ionization tracks. However, the observation at the LHC would
not constrain the NLSP lifetime significantly. Thus, we see that neutrino telescopes are
complementary to collider searches. For instance, the observation of NLSP events 
at the LHC, coupled to the non-observation in neutrino telescopes 
would point to $\sqrt{F}<10^{7}$~GeV.

Future upgrades of ICECUBE, as well as of the
water detectors ANTARES~\cite{antares} and NESTOR~\cite{nestor} will result in even better
sensitivity. As an example, in Table~\ref{tab:icetop} we show rates for an expanded version of
ICECUBE~\cite{iceplus}.

In the present letter we focused on supersymmetry.  However, many other theories give rise to
relatively long lived charged particles which can be observed by neutrino
telescopes~\cite{abc2}.
\begin{table}
\begin{center}
\caption{Number of events for extended ICECUBE~\cite{iceplus} per year assuming $\nu$ flux
is given by the WB limit. The $\tilde\ell_L$ and squark masses and the number of muons 
are as in Table~\ref{table}.}
\begin{ruledtabular}
\begin{tabular}{l|cccc}
 & $\mu$ & $m_{\tilde q}=300$~GeV & $600$~GeV & $900$~GeV \\
\hline
1 ring, 300 m & 110 & 5 & 2 & 1 \\
1 ring, 1000 m & 110 & 6 & 2 & 1 \\
4 rings, 300 m & 131 & 9 & 3 & 1 \\
4 rings, 1000 m & 140 & 16 & 5 & 2 
\label{tab:icetop}
\end{tabular}
\end{ruledtabular}
\end{center}
\end{table}

{\it Acknowledgments ---} 
We would like to thank Azriel Goldschmidt and Ian Hinchliffe for useful conversations.  
I.~A. is supported by NSF Grant Physics/Polar Programs 0071886. G.~B. and
Z.~C. are supported by the Director, Office of Science, Office of High Energy and Nuclear Physics,
of the U.~S. Department of Energy under Contract DE-AC03-76SF00098 and by the N.S.F.  under
PHY-00-98840.

\end{document}